\documentclass[thmsa,11pt]{article}
\usepackage{amsfonts}
\usepackage{amsmath}
\usepackage{amssymb}
\usepackage{sw20lart}

\input{tcilatex}
\begin{document}

\title{Biorthonormal Matrix-Product-State Analysis for Non-Hermitian
Transfer-Matrix Renormalization-Group in the Thermodynamic Limit}
\date{}
\author{Yu-Kun Huang$^{\thanks{%
\noindent Corresponding author. E-mail address: ykln@mail.njtc.edu.tw;}}$\  
\\
Graduate school of Engineering Science and Technology, Nan Jeon Institute \\
of Technology, Tainan 73746, Taiwan}
\maketitle

\begin{abstract}
We give a thorough Biorthonormal Matrix-Product-State (BMPS) analysis of the
Transfer-Matrix Renormalization-Group (TMRG) for non-Hermitian matrices in
the thermodynamic limit. The BMPS is built on a dual series of reduced
biorthonormal bases for the left and right Perron states of a non-Hermitian
matrix. We propose two alternative infinite-size Biorthonormal TMRG (iBTMRG)
algorithms and compare their numerical performance in both finite and
infinite systems. We show that both iBTMRGs produce a dual infinite-BMPS
(iBMPS) which are translationally invariant in the thermodynamic limit. We
also develop an efficient wave function transformation of the iBTMRG, an
analogy of McCulloch in the infinite-DMRG [arXiv:0804.2509 (2008)], to
predict the wave function as the lattice size is increased. The resulting
iBMPS allows for probing bulk properties of the system in the thermodynamic
limit without boundary effects and allows for reducing the computational
cost to be independent of the lattice size, which are illustrated by
calculating the magnetization as a function of the temperature and the
critical spin-spin correlation in the thermodynamic limit for a 2D classical
Ising model.

\textbf{Keywords}: Transfer-matrix renormalization-group, Density-matrix
renormalization-group, Biorthonormal matrix-product-state, Correlation
function.

\noindent \textbf{\noindent \noindent }\noindent \textbf{PACS classification}%
: 05.10.Cc; 05.50.+q; 02.70.-c; 05.70.-a
\end{abstract}

\section{Introduction}

It has been widely understood that the Matrix-Product-State (MPS) ansatz
constitutes the basis of many numerical algorithms in computational physics,
notably the Density-Matrix Renormalization-Group (DMRG) [1] and the
Time-Evolving Block Decimation (TEBD) [2]. There are two main variants of
the DMRG algorithm. The finite-size algorithm was by now realized to compute
variationally the ground state of a strongly correlated 1D quantum system
within the class of MPS [3]. While the infinite-size algorithm grows the
system by adding iteratively one or more sites and reaches asymptotically,
at the fixed point, a translationally invariant MPS state (invariant under
translations of some fixed number of lattice sites), it was originally
conceived as only for obtaining the initial MPS wave function for the
finite-size algorithm. Until recently, interest in directly obtaining the
translationally invariant MPS wave function in the thermodynamic limit was
rekindled by the ideas of infinite-TEBD (iTEBD) of Vidal [4] and
infinite-DMRG (iDMRG) of McCulloch [5]. Both approaches allow for probing
bulk properties of the system in the thermodynamic limit without the
influence of boundary conditions and the resulting infinite-MPS (iMPS) can
reduce the computational cost of real-time simulations to be independent of
the lattice size. The main difference between these two methods lies in the
scheme for the local update of the tensors in the MPS. The iTEBD applies a
single bond evolution operator to each site simultaneously which amounts to
a power method while the iDMRG obtains the center matrix variationally by
using a very efficient local eigensolver.

On the other hand, after nearly two decades of development, the application
of DMRG methods has spread over a great variety of fields. One major branch
of the DMRG is the Transfer-Matrix Renormalization-Group (TMRG) which can be
applied to, e.g., the strongly correlated classical systems [6], the
thermodynamics of 1D quantum systems at finite temperature [7], the
stochastic transfer matrix of a cellular automaton [8], the non-equilibrium
systems in statistical physics [9], and the general Markov random field in
image modeling [10]. Different from the DMRG, the TMRG usually deals with
non-Hermitian matrices which involve much more numerical demand in view of
the existence of distinct left and right Perron states (i.e., the
eigenstates associated with the maximum eigenvalue which we will refer to as
the Perron root) of the transfer matrix. This leads to the selection of the
reduced density matrix becoming ambiguous. Enss and Schollw\"{o}ck [8] had
provided a comparative discussion on several choices of the density matrix
proposed in the literature for the non-Hermitian TMRG. Unfortunately, all
these selections of density matrices cause the conventional TMRG fail to fit
the framework of MPS analysis. In a recent paper [11], the author proposed a
new TMRG algorithm called Biorthonormal Transfer-Matrix
Renormalization-Group (BTMRG) which employed a dual set of biorthonormal
bases to construct the renormalized transfer matrix and reduced the
numerical complexity for non-Hermitian matrices to be the same as the
Hermitian case. Numerical simulations for a real non-Hermitian matrix showed
that the BTMRG exhibits significant improvement on the efficiency and
accuracy than conventional TMRG. Here, a dual biorthonormal bases indicate
any two sets of vectors $\{ \left \vert \alpha \right \rangle \}_{\alpha
=1,\cdots ,m}$ and $\left \{ \left \vert \beta \right \rangle \right \}
_{\beta =1,\cdots ,m}$ that satisfy $\left \langle \alpha \right \vert \beta
\rangle =\delta _{\alpha \beta }$. In the previous paper, the BTMRG was
described in a traditional formulation. In this paper, we will show that the
employment of the biorthonormal bases enables the BTMRG to be perfectly
reformulated within the framework of MPS analysis where the two Perron
states can be represented as a dual Biorthonormal Matrix-Product-States
(BMPS). We will propose two alternative methods for building the BMPS. At
the same time of doing so, two natural questions arise: Does the
infinite-size variant of the BTMRG (iBTMRG) create asymptotically a dual
infinite biorthonormal MPS (iBMPS) that are translationally invariant? Does
there exist an efficient transformation in iBTMRG, just as the one in iDMRG
[5], for predicting the wave function as the lattice size is increased? In
this paper, both questions will be addressed confirmatively. Here, instead
of the two-site iDMRG where the non-zero truncation of the wave function
often leads to undesirable effects of the convergence of the iMPS [5], we
will use the \textquotedblleft single-site\textquotedblright \ iBTMRG scheme
which achieves zero truncation similar to the single-site iDMRG. In
addition, by using a special E$\bullet $S$\bullet $E configuration, the
critical two-point correlation functions in the thermodynamic limit,
especially when the distance between the two points approaches infinity,
will be shown to be easily obtained.

The rest of this paper is organized as follows. In section 2, we present a
brief review of the MPS analysis of the standard DMRG in finite systems and
reformulate the BTMRG algorithm with the E$\bullet $S$\bullet $E
configuration proposed in [11] entirely in terms of the BMPS language. In
section 3, we propose two alternative methods to build the BMPS of the left
and right Perron state for non-Hermitian matrices and develop an iBTMRG
algorithm to obtain the asymptotics of the dual iBMPS wave functions that
are translationally invariant. We also develop an efficient wave function
transformation in iBTMRG for predicting the BMPS as the lattice size is
increased. Section 4 compares the numerical performance of our iBTMRG
algorithms in finite systems for a real non-Hermitian matrix of an
anisotropic Ising model. The magnetization as a function of the temperature
and the critical two-point correlation function of an isotropic Ising model
in the thermodynamic limit are also plotted. Finally, in section 5, some
conclusions are drawn.

\section{Biorthonormal MPS analysis of BTMRG in finite systems}

The connection between DMRG and MPS was first found by \"{O}stlund and
Rommer [12] who identified the thermodynamic limit of DMRG with a
position-independent matrix product wave function. The discovery of the MPS
on which the DMRG operates has placed the algorithm on a firm footing,
provided a deeper understanding, and allowed a concrete and easy to
manipulate description of the DMRG. The standard DMRG and MPS formulation
was established on a series of reduced orthonormal bases. However, when
considering both left and right Perron states of non-Hermitian transfer
matrices, the MPS must be built on a dual series of reduced biorthonormal
bases which we will refer to as biorthonormal matrix-product states (BMPS).
In this section, we briefly review the finite-size BTMRG algorithm proposed
in [11] but reformulate it entirely in terms of a BMPS language. For further
information about MPS, see [13].

\subsection{MPS formulation of DMRG in finite systems}

Let us start with the MPS formulation of the standard DMRG in finite
systems. Throughout this paper, we focus on the MPS of open boundary
conditions. A spin chain can be bi-partitioned into two parts: the system
block where the spins are labeled by $s_{i}$, and the environment block
where the spins are labeled by $\varepsilon _{i}$. Then we denote an MPS on
an L-site lattice by the form

\begin{equation}
\left \vert \psi \right \rangle =\underset{s_{i},\varepsilon _{i}}{\sum }%
A_{1}^{s_{1}}\cdots A_{p}^{s_{p}}\mathit{\Lambda }E_{q}^{\varepsilon
_{q}}\cdots E_{1}^{\varepsilon _{1}}\left \vert s_{1}\cdots s_{p}\varepsilon
_{q}\cdots \varepsilon _{1}\right \rangle
\end{equation}%
where $p+q=L$ and each matrix at each site has dimension $m\times m$ with
the exception that the end matrices $A_{1}^{s_{1}}$ and $E_{1}^{\varepsilon
_{1}}$ are a row vector and a column vector respectively. The $A$-matrices ($%
E$-matrices) satisfy the left (right)-normalization condition $\underset{%
s_{i}}{\sum }A_{i}^{s_{i}\dag }A_{i}^{s_{i}}=I$ ($\underset{\varepsilon _{i}}%
{\sum }E_{i}^{\varepsilon _{i}}E_{i}^{\varepsilon _{i}\dag }=I$). It is very
useful to express an MPS in a tensor network representation (see Schollw\"{o}%
ck [13]) as in Fig. 1(a). For example, the set of matrices $A_{p}^{s_{p}}$
at the $p$-th site of the system block corresponds to a tensor $%
(A_{p})_{\alpha _{p-1},\alpha _{p}}^{s_{p}}$ with two bond indices $\alpha
_{p-1}$\ and $\alpha _{p}$ and one physical index $s_{p}$. The bond index
labels the reduced orthonormal basis state of the system block at different
lengths and the physical index labels the state of the spin. Thus the center
matrix $\mathit{\Lambda }$ represents in fact the wave function of the
superblock in the reduced orthonormal basis $\{ \left \vert \alpha _{p}\xi
_{q}\right \rangle \}_{\alpha ,\xi =1,\cdots ,m}$ (see Fig. 1(a)). The
essential notion of the MPS formulation of the finite-size variant of DMRG
is that of \textit{local updates}; that is, we free a matrix, say $%
E_{q}^{s_{q}}$, at a time while keeping the others fixed, update it by $%
A_{p+1}^{s_{p+1}}$ through an efficient optimization of the total energy,
and shift the center matrix to the right by one site. When all matrices in
the MPS are updated once, it is called a sweep. Repeat the sweep until
convergence is reached.

\subsection{E$\bullet $S$\bullet $E scheme and the related BMPS
representation}

Now consider a special bi-partitioning of the chain as indicated by the
notation E$\bullet $S$\bullet $E where the system block is taken as a
consecutive segment of sites around the center of the chain and the
environment block is taken as two equal separate segments surrounding the
system block. The superblock is updated at a time by freeing (or adding) two
spins in between the two sub-systems so that the matrices of the MPS must be
associated with two distant sites. This special configuration is shown to be
particularly adapted to the calculation of two-point correlation functions
of 1D quantum systems or 2D classical lattice models [11]. Actually, the MPS
for this E$\bullet $S$\bullet $E scheme can be formulated as equivalently as
the MPS for the standard DMRG. By folding the spin chain from the center so
that the two sub-systems and the two free spins are aligned, every two
aligned spins can be regarded as a single big spin and the resulting new
chain has the same configuration as the more standard S$\bullet $E scheme.
To avoid the notation from getting too involved, it is convenient and
comprehensive to represent this MPS as a tensor network as in Fig. 1(b)
where the tensor $(A_{p})_{\alpha _{p-1},\alpha _{p}}^{\mathbf{:}}$ , now,
with two physical indices representing the states of the two spins.
Alternatively, following McCulloch [5], such an MPS can be simply expressed
as

\begin{equation}
\left \vert \psi \right \rangle =A_{1}^{\mathbf{:}}\cdots A_{p}^{\mathbf{:}}%
\mathit{\Lambda }E_{q}^{\mathbf{:}}\cdots E_{1}^{\mathbf{:}}
\end{equation}%
In this notation, although the ket basis vectors of the MPS are suppressed
and the alphabets labeling the state of the aligned spin pairs are simply
denoted by two dots, all information are preserved. It is easy to understand
that, here, the MPS matrices are local state valued albeit not explicitly
written.

When a non-Hermitian transfer matrix $T$ is considered, the left and right
Perron state are generally distinct. Suppose the left $\left \vert \psi
\right \rangle $\ and right $\left \vert \varphi \right \rangle $ Perron
state have MPS as follows.

\begin{equation}
\left \{ 
\begin{array}{c}
\left \vert \psi \right \rangle =\overline{A}_{1}^{\mathbf{:}}\cdots 
\overline{A}_{p}^{\mathbf{:}}\overline{\mathit{\Lambda }}\overline{E}_{q}^{%
\mathbf{:}}\cdots \overline{E}_{1}^{\mathbf{:}} \\ 
\left \vert \varphi \right \rangle =\overline{B}_{1}^{\mathbf{:}}\cdots 
\overline{B}_{p}^{\mathbf{:}}\overline{\mathit{\Gamma }}\overline{F}_{q}^{%
\mathbf{:}}\cdots \overline{F}_{1}^{\mathbf{:}}%
\end{array}%
\right.
\end{equation}%
When we free the two spins associated with the matrices $\overline{E}_{q}^{%
\mathbf{:}}$ and $\overline{F}_{q}^{\mathbf{:}}$, we are actually carrying
out the local updating by maximizing the quantity $\left \langle \psi
\right
\vert T\left \vert \varphi \right \rangle -\lambda \left \langle
\psi \right
\vert \varphi \rangle $. Hence, if we impose the so-called
left- and right-biorthonormal conditions on the MPS matrices

\begin{equation}
\underset{\mathbf{:}}{\sum }\overline{B}_{i}^{\mathbf{:\dag }}\overline{A}%
_{i}^{\mathbf{:}}=\underset{\mathbf{:}}{\sum }\overline{E}_{i}^{\mathbf{:}}%
\overline{F}_{i}^{\mathbf{:\dag }}=I
\end{equation}%
then the optimization will be equivalent to an eigenvalue problem of the
reduced transfer matrix $\overline{T}$ that can be expressed as a tensor
network as in Fig. 2 (see Schollw\"{o}ck [13]) where the transfer matrix $T$
is expressed as a Matrix-Product-Operator (MPO) and $\sigma _{L}$ and $%
\sigma _{R}$ label the states of the free distant spin pair associated with
the matrices $\overline{E}_{q}^{\mathbf{:}}$ and $\overline{F}_{q}^{\mathbf{:%
}}$. The matrix elements of the reduced transfer matrix can be written
explicitly as $\overline{T}_{\alpha _{p}\sigma _{L}\sigma _{R}\xi
_{q-1},\beta _{p}\sigma _{L}^{\prime }\sigma _{R}^{\prime }\zeta
_{q-1}}=\left \langle \alpha _{p}\sigma _{L}\sigma _{R}\xi
_{q-1}\right
\vert T\left \vert \beta _{p}\sigma _{L}^{\prime }\sigma
_{R}^{\prime }\zeta _{q-1}\right \rangle $. If the pair of MPS states Eq.
(3) satisfies the biorthonormal conditions Eq. (4), we refer to the MPS as a
dual BMPS. Thus, the BTMRG algorithm can be perfectly fitted into the
framework of MPS analysis and enjoy the same numerical complexity as the
Hermitian case in DMRG algorithm. In this paper, we will propose two
alternative methods for determining the BMPS for the Perron state and
compare their numerical performance. The details will be described in
section 3.

\subsection{Density matrix and canonical form of BMPS representations}

Unlike the Hermitian Hamiltonian in DMRG that there is a unique normalized
ground state, there are two distinct Perron states in TMRG so that the
density operator for the Perron state of the system must be taken as $%
\widehat{\rho }=\left \vert \varphi \right \rangle \left \langle \psi
\right
\vert $\ (with proper normalization $\left \langle \psi \right \vert
\varphi \rangle =1$). Thus, from Eq. (3), we can readily obtain the reduced
density operator for the system and environment block as

\begin{equation}
\left \{ 
\begin{array}{c}
\widehat{\rho }_{S}=\underset{\beta _{p},\alpha _{p}}{\sum }(\overline{%
\mathit{\Gamma }}\overline{\mathit{\Lambda }}^{\dag })_{\beta _{p},\alpha
_{p}}\left \vert \beta _{p}\right \rangle \left \langle \alpha _{p}\right
\vert \\ 
\widehat{\rho }_{E}=\underset{\zeta _{q},\xi _{q}}{\sum }(\overline{\mathit{%
\Gamma }}^{\dag }\overline{\mathit{\Lambda }})_{\zeta _{q},\xi _{q}}\left
\vert \zeta _{q}\right \rangle \left \langle \xi _{q}\right \vert%
\end{array}%
\right.
\end{equation}%
Note that\ $\left \langle \alpha _{p}\right \vert \beta _{p}\rangle =\delta
_{\alpha \beta }$ and\ $\left \langle \xi _{q}\right \vert \zeta _{q}\rangle
=\delta _{\xi \zeta }$\ in view of the biorthonormal conditions. However,
given a BMPS as in Eq. (3), we can always construct another BMPS by
introducing an arbitrary invertible (non-unitary) transformation to the
matrices $\overline{E}_{q}^{\mathbf{:}}$ and $\overline{F}_{q}^{\mathbf{:}}$
(i.e., applying a non-unitary basis transformation $X$ to the current
biorthonormal bases $\{ \left \vert \xi _{q}\right \rangle \}$ and $\{
\left
\vert \zeta _{q}\right \rangle \}$) such that $\overline{E^{\prime }}%
_{q}^{\mathbf{:}}=X^{-1}\overline{E}_{q}^{\mathbf{:}}$\ and $\overline{%
F^{\prime }}_{q}^{\mathbf{:}}=X^{\dag }\overline{F}_{q}^{\mathbf{:}}$ remain
right-biorthonormal. Similar results $\overline{A^{\prime }}_{p}^{\mathbf{:}%
}=\overline{A}_{p}^{\mathbf{:}}Y^{-1\dag }$ and $\overline{B^{\prime }}_{p}^{%
\mathbf{:}}=\overline{B}_{p}^{\mathbf{:}}Y$ are obtained by applying a
transformation $Y$ to the current biorthonormal bases $\{ \left \vert \alpha
_{p}\right \rangle \}$ and $\{ \left \vert \beta _{p}\right \rangle \}$.
Thus the varied BMPS turns out to be

\begin{equation}
\left \{ 
\begin{array}{c}
\left \vert \psi \right \rangle =\overline{A}_{1}^{\mathbf{:}}\cdots 
\overline{A^{\prime }}_{p}^{\mathbf{:}}\overline{\mathit{\Lambda }^{\prime }}%
\overline{E^{\prime }}_{q}^{\mathbf{:}}\cdots \overline{E}_{1}^{\mathbf{:}}
\\ 
\left \vert \varphi \right \rangle =\overline{B}_{1}^{\mathbf{:}}\cdots 
\overline{B^{\prime }}_{p}^{\mathbf{:}}\overline{\mathit{\Gamma }^{\prime }}%
\overline{F^{\prime }}_{q}^{\mathbf{:}}\cdots \overline{F}_{1}^{\mathbf{:}}%
\end{array}%
\right.
\end{equation}%
where the new center matrices become $\overline{\mathit{\Lambda }^{\prime }}%
=Y^{\dag }\overline{\mathit{\Lambda }}X$ and $\overline{\mathit{\Gamma }%
^{\prime }}=Y^{-1}\overline{\mathit{\Gamma }}X^{-1\dag }$ and the new
reduced density operators are simply a similar transform of the old density
operators: $\widehat{\rho }_{S}^{\prime }=Y^{-1}\widehat{\rho }_{S}Y$\ and $%
\widehat{\rho }_{E}^{\prime }=X^{-1}\widehat{\rho }_{E}X$. This implies that
if we employ a suitable transformation, we can canonize the form of the BMPS
representation.

Given the BMPS in Eq. (3), assume the density matrix $\rho _{S}=\overline{%
\mathit{\Gamma }}\overline{\mathit{\Lambda }}^{\dag }=YDY^{-1}$ is
diagonalizable, then it can be readily obtained that $\rho _{E}=\overline{%
\mathit{\Gamma }}^{\dag }\overline{\mathit{\Lambda }}=XDX^{-1}$ where $X=%
\overline{\mathit{\Gamma }}^{\dag }Y^{-1\dag }$. Since $X$ and $Y$ are
unique up to a scaling, we can replace $X$ and $Y$ by $X\mathit{\lambda }%
_{1} $ and $Y\mathit{\lambda }_{2}$\ where $\mathit{\lambda }_{1}$ and $%
\mathit{\lambda }_{2}$ are two diagonal matrices. Accordingly, we have new
center matrices $\overline{\mathit{\Lambda }^{\prime }}=\mathit{\lambda }%
_{2}D\mathit{\lambda }_{1}=$ $\sqrt{D}$ and $\overline{\mathit{\Gamma }%
^{\prime }}=\mathit{\lambda }_{2}^{-1}\mathit{\lambda }_{1}^{-1}=$ $\sqrt{D}$
by choosing $\mathit{\lambda }_{2}=$ $D^{-1/2}\mathit{\lambda }_{1}^{-1}$.
Thus the canonical form of BMPS reads

\begin{equation}
\left \{ 
\begin{array}{c}
\left \vert \psi \right \rangle =\overline{A}_{1}^{\mathbf{:}}\cdots 
\overline{A}_{p}^{\mathbf{:}}\sqrt{D}\overline{E}_{q}^{\mathbf{:}}\cdots 
\overline{E}_{1}^{\mathbf{:}} \\ 
\left \vert \varphi \right \rangle =\overline{B}_{1}^{\mathbf{:}}\cdots 
\overline{B}_{p}^{\mathbf{:}}\sqrt{D}\overline{F}_{q}^{\mathbf{:}}\cdots 
\overline{F}_{1}^{\mathbf{:}}%
\end{array}%
\right.
\end{equation}%
and the density matrix bears the canonical form $\rho _{S}=\rho _{E}=D$.
Surprisingly, according to our practical simulation, for large enough system
size, $\rho _{S}=YDY^{-1}$\ always exits and the eigenvalues are always real
non-negative.

\section{Biorthonormal MPS analysis of iBTMRG in the thermodynamic limit}

The iDMRG grows the system at a time by adding one or two sites at the
center of the lattice and produce a fixed point of iMPS that is
translationally invariant. By thinking each aligned spin pairs as a big
spin, our iBTMRG with the E$\bullet $S$\bullet $E configuration can be
formulated as equivalently as the iDMRG. Throughout this paper, we choose
the \textquotedblleft single-site\textquotedblright \ iBTMRG algorithm based
on the following reasons. First, \textquotedblleft
two-site\textquotedblright \ iBTMRG means adding four spins at a time where
the dimension of the reduced transfer matrix significantly increases.
Second, the inevitable non-zero truncation of the wave function of the
two-site iBTMRG often leads to a less-well converged wave function [5].
Third, the single-site iBTMRG can achieve zero truncation of the wave
function and the problem of being trapped in local minimum can be avoided by
introducing White's correction [14] to the reduced biorthonormal bases.

In this paper, many formulations are derived from a common basic procedure
which we will refer to as the \textit{Biorthonormalization Procedure}. Given
two arbitrary bases $\{ \left \vert \alpha \right \rangle \}_{\alpha
=1,\cdots ,m}$ and $\left \{ \left \vert \beta \right \rangle \right \}
_{\beta =1,\cdots ,m}$ , suppose $A\equiv \left[ \left \vert \alpha
\right
\rangle _{\alpha =1,\cdots ,m}\right] $ and $B\equiv \left[
\left
\vert \beta \right \rangle _{\beta =1,\cdots ,m}\right] $\ are two
matrices formed with $\left \vert \alpha \right \rangle $ and $\left \vert
\beta \right
\rangle $ as their columns respectively. By carrying out the
Singular-Value-Decomposition (SVD) $A^{\dag }B\equiv U\mathit{\Sigma }%
V^{\dag }$, we can readily obtain $(AP)^{\dag }(BW)=I$ where $P=U\mathit{%
\Sigma }^{-1/2}$\ and $W=V\mathit{\Sigma }^{-1/2}$. This means that, by
applying the non-unitary basis transformation $P$ and $W$ to the original
bases $\{ \left \vert \alpha \right \rangle \}$ and $\left \{ \left \vert
\beta \right \rangle \right \} $, we can obtain a dual biorthonormal bases $%
AP$ and $BW$.

\subsection{Two alternative iBTMRG algorithms in the thermodynamic limit}

We first propose an iBTMRG algorithm where the main steps are depicted in
Fig. 3. The iBTMRG starts with an initialization procedure (i.e., $n=1$)
which selects a minimum length of the lattice (depends on the number $m$ of
states kept by BTMRG), obtains the respective reduced bases of the two
sub-systems from the Perron states, and proceeds to biorthonormalize them.
The routine in Fig. 3 grows the lattice by adding first two sites to the
system block and then adding two sites to the environment block, which
constitute a period of the whole algorithm with respect to a repeated
fragment of the final iBMPS. At each time when adding new sites to the
lattice, we obtain the Perron states (e.g., $\psi ^{:}$) in the reduced
biorthornormal bases, carry out an SVD for the Perron states to obtain the
reduced bases of the enlarged block (e.g., $\psi ^{:}=A_{n+1}^{\mathbf{:}}%
\mathit{\Lambda }_{n+1}$ where $\underset{\mathbf{:}}{\sum }A_{n+1}^{\mathbf{%
:\dag }}A_{n+1}^{\mathbf{:}}=I$), and biorthonormalize them to build a new
biorthonormal bases of the enlarged block (e.g., the right bond states of
the tensors $\overline{A}_{n+1}^{\mathbf{:}}$ and $\overline{B}_{n+1}^{%
\mathbf{:}}$). If the lattice size grows large enough, we can additionally
canonize the BMPS. This step is convenient for the purpose of checking the
convergence of the algorithm. According to the previous description, the
biorthonormalization and the canonization procedures are simply to apply two
successive non-unitary basis transformation to the bases obtained from the
SVD, we can combine these transformation together and denote them by $P$, $W$%
, $Q$, and $R$ with respect to $A$, $B$, $E$, and $F$ matrices respectively
(e.g., $\overline{A}_{n+1}^{\mathbf{:}}=A_{n+1}^{\mathbf{:}}P_{n+1}$\ and $%
\overline{E}_{n+1}^{\mathbf{:}}=Q_{n+1}E_{n+1}^{\mathbf{:}}$, and thus $%
\overline{\mathit{\Lambda }}_{n+1}=P_{n+1}^{-1}\mathit{\Lambda }%
_{n+1}Q_{n+1}^{-1}$). Note that, during the SVD in step 2, we keep zero
truncation of the wave function when obtaining the reduced basis of the
enlarged block. The resulting BMPS read

\begin{equation}
\left \{ 
\begin{array}{c}
\left \vert \psi \right \rangle =\overline{A}_{1}^{\mathbf{:}}\cdots 
\overline{A}_{n}^{\mathbf{:}}\overline{\mathit{\Lambda }}_{n}\overline{E}%
_{n}^{\mathbf{:}}\cdots \overline{E}_{1}^{\mathbf{:}} \\ 
\left \vert \varphi \right \rangle =\overline{B}_{1}^{\mathbf{:}}\cdots 
\overline{B}_{n}^{\mathbf{:}}\overline{\mathit{\Gamma }}_{n}\overline{F}%
_{n}^{\mathbf{:}}\cdots \overline{F}_{1}^{\mathbf{:}}%
\end{array}%
\right.
\end{equation}%
The above algorithm builds the biorthonormalized bases from the reduced
bases derived from the SVD that are based on another biorthonormal bases.
One may notice that there exist actually a lot of possibilities to build the
biorthonormal bases. Nevertheless, the process of SVD is critical. We can
also jump the SVD and directly biorthonormalize $\psi ^{:}$ and $\varphi
^{:} $. But such a numerical procedure will become unstable very quickly.
Here, we provide another algorithm which is very different from the above
algorithm in nature. The alternative algorithm is depicted in Fig. 3 with
the steps 2-3 enclosed in a rectangle being replaced by the routine depicted
in Fig. 4. To the contrary, this algorithm constructs the biorthonormalized
bases from the reduced bases derived from the SVD that are based on two 
\textit{orthonormal} bases. The first major change lies in step 2 where
before performing SVD the Perron states must be restored to be in terms of
their own orthonormal bases, e.g., $\psi ^{\prime :}=P_{n}\psi ^{:}Q_{n}$.
The second change lies in step 3 where before performing the
biorthonormalization procedure the reduced bases derived from the SVD must
be transformed back to be in terms of the previous biorthonormal bases,
e.g., $A_{n+1}^{\prime \mathbf{:}}=P_{n}^{-1}A_{n+1}^{\mathbf{:}}$. As it
turns out, two remarkable properties emerge, e.g., $\overline{A}_{n+1}^{%
\mathbf{:}}=P_{n}^{-1}A_{n+1}^{\mathbf{:}}P_{n+1}$\ and $\overline{A}_{1}^{%
\mathbf{:}}\cdots \overline{A}_{n}^{\mathbf{:}}=A_{1}^{\mathbf{:}}\cdots
A_{n}^{\mathbf{:}}P_{n+1}$. Similar relation holds for the $B$, $E$, and $F$
matrices. Accordingly, the BMPS can also be expressed in terms of two series
of reduced orthonormal bases.

\begin{equation}
\left \{ 
\begin{array}{c}
\left \vert \psi \right \rangle =A_{1}^{\mathbf{:}}\cdots A_{n}^{\mathbf{:}}%
\mathit{\Lambda }_{n}E_{n}^{\mathbf{:}}\cdots E_{1}^{\mathbf{:}} \\ 
\left \vert \varphi \right \rangle =B_{1}^{\mathbf{:}}\cdots B_{n}^{\mathbf{:%
}}\mathit{\Gamma }_{n}F_{n}^{\mathbf{:}}\cdots F_{1}^{\mathbf{:}}%
\end{array}%
\right.
\end{equation}%
where $\mathit{\Lambda }_{n}=P_{n}\overline{\mathit{\Lambda }}_{n}Q_{n}$ and 
$\mathit{\Gamma }_{n}=W_{n}\overline{\mathit{\Gamma }}_{n}R_{n}$. For
convenience, in this paper, the former algorithm will be referred to as 
\textbf{iBTMRG A} and the latter \textbf{iBTMRG B}.

\subsection{Efficient wave function prediction and the fixed point of the
iBMPS}

In both algorithms, the iBTMRG uses a local eigensolver to obtain the Perron
vectors. A good initial guess for the eigenvector will significantly improve
the performance of the eigensolver. In [5], McCulloch has developed a wave
function transformation in iDMRG for the prediction of the wave function as
the lattice size is increased. Such a transformation can be effectively
translated to our iBTMRG. Therefore, at the end of the first half-period of
the process in Fig. 3, the prediction of the next Perron vectors will be

\begin{equation}
\left \{ 
\begin{array}{c}
\psi _{trial}^{\mathbf{:}}=D_{n+1}^{1/2}D_{n}^{-1/2}\overline{A}_{n+1}^{%
\mathbf{:}}\overline{\mathit{\Lambda }}_{n+1}=D_{n+1}^{1/2}D_{n}^{-1/2}\psi
^{\mathbf{:}} \\ 
\varphi _{trial}^{\mathbf{:}}=D_{n+1}^{1/2}D_{n}^{-1/2}\overline{B}_{n+1}^{%
\mathbf{:}}\overline{\mathit{\Gamma }}_{n+1}=D_{n+1}^{1/2}D_{n}^{-1/2}%
\varphi ^{\mathbf{:}}%
\end{array}%
\right.
\end{equation}%
where $D_{n+1}^{1/2}$ and $\psi ^{\mathbf{:}}$ are the current canonical
center matrix and the current Perron vector. Similar results go to the
Perron vectors prediction at the end of the second half-period in Fig. 3.

\begin{equation}
\left \{ 
\begin{array}{c}
\psi _{trial}^{\mathbf{:}}=\overline{\mathit{\Lambda }}_{n+1}\overline{E}%
_{n+1}^{\mathbf{:}}D_{n}^{-1/2}D_{n+1}^{1/2}=\psi ^{\mathbf{:}%
}D_{n}^{-1/2}D_{n+1}^{1/2} \\ 
\varphi _{trial}^{\mathbf{:}}=\overline{\mathit{\Gamma }}_{n+1}\overline{F}%
_{n+1}^{\mathbf{:}}D_{n}^{-1/2}D_{n+1}^{1/2}=\varphi ^{\mathbf{:}%
}D_{n}^{-1/2}D_{n+1}^{1/2}%
\end{array}%
\right.
\end{equation}%
Note that, for finite fixed $n$, the center matrix $D_{n}^{1/2}$ may not be
identical for the two half-periods in Fig. 3 so that the center matrices in
Eqs. (10-11) should use their own $D_{n}^{1/2}$. When the size of the
lattice is small, the $D_{n}^{1/2}$ can be replace by $\overline{\mathit{%
\Lambda }}_{n}$ and $\overline{\mathit{\Gamma }}_{n}$.

At the fixed point of the iBTMRG, the translational invariant iBMPS should
bear the canonical form

\begin{equation}
\left \{ 
\begin{array}{c}
\left \vert \psi \right \rangle =\cdots (D_{n-1}^{-1/2}\overline{A}_{n}^{%
\mathbf{:}}D_{n}^{1/2}\overline{E}_{n}^{\mathbf{:}})(D_{n-1}^{-1/2}\overline{%
A}_{n}^{\mathbf{:}}D_{n}^{1/2}\overline{E}_{n}^{\mathbf{:}})\cdots \\ 
\left \vert \varphi \right \rangle =\cdots (D_{n-1}^{-1/2}\overline{B}_{n}^{%
\mathbf{:}}D_{n}^{1/2}\overline{F}_{n}^{\mathbf{:}})(D_{n-1}^{-1/2}\overline{%
B}_{n}^{\mathbf{:}}D_{n}^{1/2}\overline{F}_{n}^{\mathbf{:}})\cdots%
\end{array}%
\right.
\end{equation}%
where we take $n$ to be the iteration step when the convergence criterion is
met. At the fixed point, the density matrix would be invariant no matter
where the bond is located. This leads to the fixed point criterion $\underset%
{\mathbf{:}}{\sum }\overline{A}_{n}^{\mathbf{:}}D_{n}\overline{B}_{n}^{%
\mathbf{:\dag }}=D_{n-1}$. Since $Tr(D_{n})=1$ (with proper normalization $%
\left \langle \psi ^{:}\right \vert \varphi ^{:}\rangle =1$ ), one way of
measuring the closeness of the two density matrices $D_{n-1}$ and $%
D_{n}^{\prime }=\underset{\mathbf{:}}{\sum }\overline{A}_{n}^{\mathbf{:}%
}D_{n}\overline{B}_{n}^{\mathbf{:\dag }}$ is given by the Kullback-Leibler
divergence

\begin{equation}
D_{KL}=\underset{i}{\sum }D_{n-1}(i)\log (\frac{D_{n-1}(i)}{D_{n}^{\prime
}(i)})
\end{equation}%
In this paper, Eq. (13) is utilized to check the convergence of our iBTMRG.

\subsection{Normalization of the iBMPS}

The fixed point criterion $\underset{\mathbf{:}}{\sum }\overline{A}_{n}^{%
\mathbf{:}}D_{n}\overline{B}_{n}^{\mathbf{:\dag }}=D_{n-1}$\ is equivalent
to $\underset{\mathbf{:}}{\sum }(D_{n-1}^{-1/2}\overline{A}_{n}^{\mathbf{:}%
}D_{n}^{1/2})(D_{n-1}^{-1/2}\overline{B}_{n}^{\mathbf{:}}D_{n}^{1/2})^{\dag
}=I$ which means the two matrices $D_{n-1}^{-1/2}\overline{A}_{n}^{\mathbf{:}%
}D_{n}^{1/2}$ and $D_{n-1}^{-1/2}\overline{B}_{n}^{\mathbf{:}}D_{n}^{1/2}$
satisfy the right-biorthonormal condition. Similarly, $D_{n}^{1/2}\overline{E%
}_{n}^{\mathbf{:}}D_{n-1}^{-1/2}$\ and $D_{n}^{1/2}\overline{F}_{n}^{\mathbf{%
:}}D_{n-1}^{-1/2}$ must satisfy the left-biorthonormal condition. Thus, the
overlap of the iBMPS in Eq. (12) must be exactly $\left \langle \psi
\right
\vert \varphi \rangle =1$ in the fixed point. The normalization $%
\left
\langle \psi \right \vert \varphi \rangle =1$ is important when we
want to apply the iBMPS to the calculation of expectation value or
correlation functions in the thermodynamic limit. However, in practical
simulations, the finite number $m$ of states kept by BTMRG and finite
iterations can only reach an approximation of the fixed point so that the
iBMPS will not be exactly normalized. In [15], Or\'{u}s and Vidal have
developed a method for the orthonormalization of an iMPS in iTEBD, which can
also be effectively translated to our iBTMRG. Define two transfer operator $%
T_{R}$ and $T_{L}$ as shown in Fig. 5. In order to achieve the
right-biorthonormal condition, assume $\mathit{\Omega }_{R}$ is the right
dominant eigenvector of $T_{R}$ and an SVD allows to decompose $\mathit{%
\Omega }_{R}=U_{R}V_{R}$. We now perform two similarity transformations to
the unit cells in Eq. (12) which lead to

\begin{equation}
\left \{ 
\begin{array}{c}
\left \vert \psi \right \rangle =\cdots (U_{R}^{-1}D_{n-1}^{-1/2}\overline{A}%
_{n}^{\mathbf{:}}D_{n}^{1/2}\overline{E}_{n}^{\mathbf{:}%
}U_{R})(U_{R}^{-1}D_{n-1}^{-1/2}\overline{A}_{n}^{\mathbf{:}}D_{n}^{1/2}%
\overline{E}_{n}^{\mathbf{:}}U_{R})\cdots \\ 
\left \vert \varphi \right \rangle =\cdots (V_{R}^{-1}D_{n-1}^{-1/2}%
\overline{B}_{n}^{\mathbf{:}}D_{n}^{1/2}\overline{F}_{n}^{\mathbf{:}%
}V_{R})(V_{R}^{-1}D_{n-1}^{-1/2}\overline{B}_{n}^{\mathbf{:}}D_{n}^{1/2}%
\overline{F}_{n}^{\mathbf{:}}V_{R})\cdots%
\end{array}%
\right.
\end{equation}%
Similar arguments apply to $T_{L}$ and its left dominant eigenvector $%
\mathit{\Omega }_{L}=U_{L}V_{L}$ and lead to the left-biorthonormal iBMPS

\begin{equation}
\left \{ 
\begin{array}{c}
\left \vert \psi \right \rangle =\cdots (U_{L}\overline{A}_{n}^{\mathbf{:}%
}D_{n}^{1/2}\overline{E}_{n}^{\mathbf{:}}D_{n-1}^{-1/2}U_{L}^{-1})(U_{L}%
\overline{A}_{n}^{\mathbf{:}}D_{n}^{1/2}\overline{E}_{n}^{\mathbf{:}%
}D_{n-1}^{-1/2}U_{L}^{-1})\cdots \\ 
\left \vert \varphi \right \rangle =\cdots (V_{L}\overline{B}_{n}^{\mathbf{:}%
}D_{n}^{1/2}\overline{F}_{n}^{\mathbf{:}}D_{n-1}^{-1/2}V_{L}^{-1})(V_{L}%
\overline{B}_{n}^{\mathbf{:}}D_{n}^{1/2}\overline{F}_{n}^{\mathbf{:}%
}D_{n-1}^{-1/2}V_{L}^{-1})\cdots%
\end{array}%
\right.
\end{equation}%
Accordingly, update the matrices: $\overline{A}_{n}^{\mathbf{:}}\leftarrow
U_{L}\overline{A}_{n}^{\mathbf{:}}$ , $\overline{B}_{n}^{\mathbf{:}%
}\leftarrow V_{L}\overline{B}_{n}^{\mathbf{:}}$ , $\overline{E}_{n}^{\mathbf{%
:}}\leftarrow \overline{E}_{n}^{\mathbf{:}}U_{R}$ , $\overline{F}_{n}^{%
\mathbf{:}}\leftarrow \overline{F}_{n}^{\mathbf{:}}V_{R}$ , and $%
D_{n-1}^{1/2}\leftarrow V_{L}D_{n-1}^{1/2}U_{R}$. Then the lattice can be
separated into two parts at any bond location where the iBMPS can be
expressed as similar to as Eq. (7)

\begin{equation}
\left \{ 
\begin{array}{c}
\left \vert \psi \right \rangle =\cdots (\overline{A}_{n}^{\mathbf{:}%
}D_{n}^{1/2}\overline{E}_{n}^{\mathbf{:}%
}D_{n-1}^{-1/2})D_{n-1}^{1/2}(D_{n-1}^{-1/2}\overline{A}_{n}^{\mathbf{:}%
}D_{n}^{1/2}\overline{E}_{n}^{\mathbf{:}})\cdots \\ 
\left \vert \varphi \right \rangle =\cdots (\overline{B}_{n}^{\mathbf{:}%
}D_{n}^{1/2}\overline{F}_{n}^{\mathbf{:}%
}D_{n-1}^{-1/2})D_{n-1}^{1/2}(D_{n-1}^{-1/2}\overline{B}_{n}^{\mathbf{:}%
}D_{n}^{1/2}\overline{F}_{n}^{\mathbf{:}})\cdots%
\end{array}%
\right.
\end{equation}

\section{Example: 2D classical systems}

In this section, we test our \textbf{iBTMRG A} and \textbf{iBTMRG B}
algorithms on a 2D classical Ising model. The target matrix is $\mathit{\Pi }%
_{q}^{N}$ where $\mathit{\Pi }_{q}$ is the non-Hermitian fundamental
transfer matrix of the general local energy function (LEF)-parameterized
Markov random field on an infinitely-long vertical twisted cylindrical
lattice with peripheral length $N$ [10, 16] (The Ising model is just a
special case of a Markov random field). This matrix is intimately related to
a 2D Markov additive process and enjoys a very special SVD structure and
many fascinating properties [16].

\subsection{iBTMRG in finite systems}

Our iBTMRG algorithms can also be applied to the partition function
calculation of classical models in finite lattices. The method to build the
reduced biorthonormal bases in \textbf{iBTMRG A} is essentially the same as
the previous BTMRG in [11] which we will refer to as \textbf{iBTMRG C} for
convenience. However, the system-growing strategy of \textbf{iBTMRG C} is
different from that of the single-site iBTMRG. Upon the same E$\bullet $S$%
\bullet $E configuration, the previous BTMRG grows the system by adding two
sites at each iteration to the system block while keeping the environment
block fixed and small. To our knowledge, both proposed single-site iBTMRG
algorithms had never been tested before. Here, we will compare the
performance of our iBTMRG algorithms and the previous one by computing the
free energy of the Ising model on a finite 2D lattice. Figure 6 shows the
error of the free energy (i.e., the logarithm of the Perron root of the
transfer matrix $\mathit{\Pi }_{q}$) of an anisotropic Ising model ($%
J_{x}=-J_{y}$ where $J_{x}$ and $J_{y}$ represent the interaction between
horizontally and vertically neighboring spins) at the critical temperature $%
T_{c}=2/\log (1+\sqrt{2})$ for system size $N=160$ and various number of
states $m$ kept by iBTMRG. We also compare the performance of the
finite-size variant of BTMRG (fBTMRG) for the three algorithms where the
Perron states from the iBTMRG were used as the initial BMPS states and were
further optimized variationally. From Fig. 6, we can see that \textbf{iBTMRG
A} and \textbf{B} exhibit almost identical performance which were further
slightly improved by their variational finite-size variants. For fBTMRGs,
algorithms A and B give rise to very close results and only outperform
algorithm C a bit. Although \textbf{iBTMRG C} has the merit of high
efficiency, it is interesting to note that its accuracy is far worse and
remain constant for all values of $m$. This is because, in the
system-growing stage, the size of the environment block is kept fixed and
small so that the entanglements between the two subsystems remain fixed and
small irrespective of the reduced basis size $m$.

\subsection{iBTMRG in the thermodynamic limit}

In iBTMRG, finding the Perron states by an iterative eigensolver is the most
time-consuming part so that the overlap between the initial wave function
and the variational optimum $\left \langle \psi _{trial}^{\mathbf{:}%
}\right
\vert \psi _{opt}^{\mathbf{:}}\rangle $ will dominate the
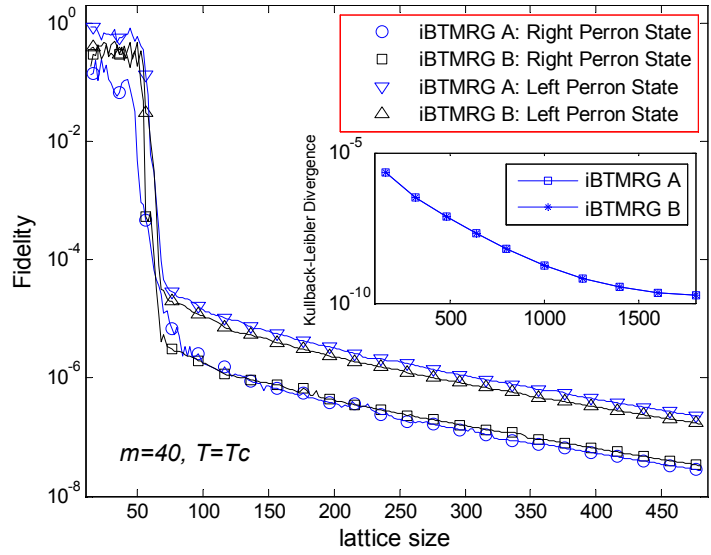
performance of an iBTMRG algorithm. Figure 7 shows the fidelity $%
1-\left
\langle \psi _{trial}^{\mathbf{:}}\right \vert \psi _{opt}^{\mathbf{%
:}}\rangle $ (see McCulloch [5]) between the Perron states predictors and
the optimal ones for the above anisotropic Ising model at criticality with $%
m=40$ states kept in the reduced basis. Before the peripheral size of the
system grows larger than $N=50$, the center matrices in Eqs. (10-11) must be
replaced by the un-canonized center matrix $\overline{\mathit{\Lambda }}_{n}$
and $\overline{\mathit{\Gamma }}_{n}$ since the density matrices cannot be
diagonalized to real positive eigenvalues until $N=50$. From Fig. 7, the
un-canonized center matrices seem inappropriate for the usage of wave
function transformation. Once the canonical center matrices are utilized,
the fidelity quickly drops down to $10^{-6}$ and continues to decay in a
nearly constant rate. Moreover, both predictions for algorithms A and B have
very close effectiveness and the right Perron state transformation appears
to prevail with respect to the left Perron state transformation. Another
important issue concerns the convergence of the translationally invariant
fixed point of the iBMPS. In this paper, we use the Kullback-Leibler
divergence to monitor the convergence of the algorithm. The inset of Fig. 7
shows the convergence to the fixed point of the iBMPS with respect to the
number of iterations. In view of the degeneracy of the Perron root and the
long-range spin correlation at the criticality, the convergence appears to
be quite slow. However, in off-critical regions, the convergence can be as
fast as reaching $10^{-12}$ as the size exceeding $N=200$.

Once we have the translationally invariant iBMPS (after normalization $%
\left
\langle \psi \right \vert \varphi \rangle =1$), the calculation of
the expectation value or the two-point correlator in the thermodynamic limit
can be very efficient. For illustration, we take an isotropic ($%
J_{x}=J_{y}=J $) Ising model as an example and calculate its magnetization
as a function of the temperature and its spin-spin correlation function at
the critical temperature. Now consider Eq. (16), if we free the unit cells $%
\psi _{\alpha ,\xi }^{\sigma _{L}\sigma _{L}^{\prime },\sigma _{R}\sigma
_{R}^{\prime }}\equiv D_{n-1}^{1/2}(D_{n-1}^{-1/2}\overline{A}_{n}^{\sigma
_{L}\sigma _{R}}D_{n}^{1/2}\overline{E}_{n}^{\sigma _{L}^{\prime }\sigma
_{R}^{\prime }})$ and $\varphi _{\beta ,\zeta }^{\sigma _{L}\sigma
_{L}^{\prime },\sigma _{R}\sigma _{R}^{\prime }}\equiv
D_{n-1}^{1/2}(D_{n-1}^{-1/2}\overline{B}_{n}^{\sigma _{L}\sigma
_{R}}D_{n}^{1/2}\overline{F}_{n}^{\sigma _{L}^{\prime }\sigma _{R}^{\prime
}})$ then we have Perron states in the thermodynamic limit expressed as $%
\psi _{\alpha ,\xi }^{\sigma _{L}\sigma _{L}^{\prime },\sigma _{R}\sigma
_{R}^{\prime }}$ and $\varphi _{\beta ,\zeta }^{\sigma _{L}\sigma
_{L}^{\prime },\sigma _{R}\sigma _{R}^{\prime }}$ with respect to a dual
biorthonormal bases. Note that the spins $\sigma _{L}\sigma _{L}^{\prime }$
and $\sigma _{R}\sigma _{R}^{\prime }$ are two neighboring spins pairs
located on the left and right side of the system block, and the spins $%
\sigma _{L}\sigma _{R}$ and $\sigma _{L}^{\prime }\sigma _{R}^{\prime }$ are
two (infinitely) far distant aligned spins. Therefore, the magnetization in
the thermodynamic limit can be calculated by

\begin{equation}
\left \langle \sigma _{R}\right \rangle =\underset{\sigma _{R}}{\sum }\sigma
_{R}\underset{\alpha =\beta ,\xi =\zeta }{\sum }\underset{\sigma _{L}\sigma
_{L}^{\prime }\sigma _{R}^{\prime }}{\sum }\psi _{\alpha ,\xi }^{\sigma
_{L}\sigma _{L}^{\prime },\sigma _{R}\sigma _{R}^{\prime }}\varphi _{\beta
,\zeta }^{\sigma _{L}\sigma _{L}^{\prime },\sigma _{R}\sigma _{R}^{\prime }}
\end{equation}%
For an infinite 2D isotropic Ising model, the exact solution for the
magnetization is $\left( 1-\left( \sinh (2J/k_{B}T)\right) ^{-4}\right)
^{1/8}$ [17]. Here, we compute the magnetization by Eq. (17) and the exact
formula as in Fig. 8 where the inset shows the relative error between the
numerical result and the exact one. Similarly, the spin-spin correlation
between two infinitely far distant spins in the thermodynamic limit can be
calculated by

\begin{equation}
\left \langle \sigma _{L}\sigma _{R}\right \rangle =\underset{\sigma
_{L}\sigma _{R}}{\sum }\sigma _{L}\sigma _{R}\underset{\alpha =\beta ,\xi
=\zeta }{\sum }\underset{\sigma _{L}^{\prime }\sigma _{R}^{\prime }}{\sum }%
\psi _{\alpha ,\xi }^{\sigma _{L}\sigma _{L}^{\prime },\sigma _{R}\sigma
_{R}^{\prime }}\varphi _{\beta ,\zeta }^{\sigma _{L}\sigma _{L}^{\prime
},\sigma _{R}\sigma _{R}^{\prime }}
\end{equation}%
At the critical temperature, the exact spin-spin correlation function scales
as $G(r)\propto r^{-1/4}\left( 1+O(r^{-2})\right) $ [18]. Ideally, Eq. (18)
will be zero but in practice it can only be seen as the correlation of two
spins separated by a distance $r$ equal to half of the system size $N$ where
the fixed point criterion is met. This implies that we can regard Eq. (18)
as the correlation function in the thermodynamic limit as $r=N/2$. Figure 9
shows the scaling behavior of the spin-spin correlator compared with the
exact solution. It is worthy of noting that, in the low-temperature region
nearby the criticality, our iBTMRG behaves quite prone to getting stuck.
Fortunately, by introducing White's correction [14] to the reduced
biorthonormalized bases, the algorithm exhibits much improved efficiency and
convergence with the correction weight around $10^{-3}$-$10^{-4}$. In [15],
Or\'{u}s and Vidal have conducted the same calculation as in Figs. 8-9 by
using the iTEBD algorithm for the same Ising model (see Figs. 11-12 in
[15]). For the magnetization, although the iTEBD has achieved a better
accuracy than iBTMRG, they are not on the same conditions. In addition to
the main difference already mentioned in the introduction, another
difference is that the transfer matrix they have dealt with is Hermitian
which is much numerically well-conditioned than the non-Hermitian case. More
importantly, for the two-point correlation calculation, instead of
evaluating a very long tensor network as in [15], we use a very efficient
formula Eq. (18) and obtain a result that is comparable with respect to the
iTEBD.

\section{Conclusions}

In this paper, we give a thorough Biorthonormal Matrix-Product-State (BMPS)
analysis of the Transfer-Matrix Renormalization-Group (TMRG) for
non-Hermitian matrices and propose a \textquotedblleft
single-site\textquotedblright \ infinite-size Biorthonormal TMRG (iBTMRG)
algorithm to directly obtain a dual infinite-BMPS (iBMPS) which are
translationally invariant in the thermodynamic limit. Different from the
standard DMRG that the MPS was established on a series of reduced
orthonormal bases, the BMPS are built on a dual series of reduced
biorthonormal bases for the left and right Perron states of a non-Hermitian
matrix. We propose two alternative methods for the construction of the dual
biorthonormal bases and compare their numerical performance in both finite
and infinite systems. We also develop an efficient wave function
transformation of the iBTMRG, an analogy of McCulloch [5] in the
infinite-DMRG, to predict the wave function as the lattice size is
increased. The resulting translationally invariant iBMPS not only allows for
investigating bulk properties of a strongly correlated system in the
thermodynamic limit without the boundary effect but also allows for the
evaluation of expectation and two-point correlation function of the system
very efficiently. For illustration, we calculate the magnetization as a
function of the temperature and the critical spin-spin correlation function
in the thermodynamic limit for a 2D classical Ising model.

\noindent {\Large References}\medskip

\noindent \lbrack 1]\textbf{\ }S. R. White, Phys. Rev. Lett. \textbf{69},
2863 (1992); I. Peschel, X. Wang, M. Kaulke and K. Hallberg (eds.) \textit{%
Density Matrix Renormalization}, Lecture Notes in Physics (Springer, Berlin,
1999); U. Schollw\"{o}ck, Rev. Mod. Phys. \textbf{77}, 259-3152005 (2005);
K. Hallberg, Adv. Phys. \textbf{55}, 477 (2006)

\noindent \lbrack 2] G. Vidal, Phys. Rev. Lett. \textbf{91}, 147902 (2003);
G. Vidal, Phys. Rev. Lett. \textbf{93}, 040502 (2004)

\noindent \lbrack 3] J. Dukelsky et al., Europhys. Lett. \textbf{43}, 457
(1998); T. Nishino et al., Int. J. Mod. Phys. B \textbf{13}, 1 (1999); F.
Verstraete, D. Porras, and J. I. Cirac, Phys. Rev. Lett. \textbf{93}, 227205
(2004); F. Verstraete, V. Murg, and J. I. Cirac, Adv. Phys. \textbf{57}, 143
(2008)

\noindent \lbrack 4] G. Vidal, Phys. Rev. Lett. \textbf{98}, 070201 (2007)

\noindent \lbrack 5] I. P. McCulloch, arXiv:0804.2509 (2008)

\noindent \lbrack 6] T. Nishino, J. Phys. Soc. Jpn. \textbf{64}, 3598
(1995); T. Nishino and K. Okunishi, J. Phys. Soc. Jpn. \textbf{66}, 3040
(1997)

\noindent \lbrack 7] R. J. Bursill, T. Xiang, and G. A. Gehring, J. Phys.:
Condens. Matter \textbf{8}, L583 (1996); X. Wang and T. Xiang, Phys. Rev. B 
\textbf{56}, 5061 (1997); N. Shibata, J. Phys. A \textbf{36}, 381 (2003); A.
E. Feiguin and S. R. White, Phys. Rev. B \textbf{72}, 220401 (2005)

\noindent \lbrack 8] A. Kemper, A. Schadschneider and J. Zittartz, J. Phys.
A: Math. Gen. \textbf{34}, L279 (2001); T. Enss and U. Schollw\"{o}ck, J.
Phys. A: Math. Gen. \textbf{34}, 7769 (2001)

\noindent \lbrack 9] Y. Hieida, J. Phys. Soc. Jpn. \textbf{67}, 369 (1998);
E. Carlon, M. Henkel and U. Schollw\"{o}ck, Phys. Rev. E \textbf{63}, 036101
(2001); S. Trebst et al., Phys. Rev. Lett. \textbf{96}, 250402 (2006); M.
Cramer et al., Phys. Rev. Lett. \textbf{101}, 063001 (2008); T. Barthel et
al., Phys. Rev. A \textbf{79}, 053627 (2009)

\noindent \noindent \lbrack 10] Y.-K. Huang and S.-N. Yu, Physica A \textbf{%
390}, 801 (2011)

\noindent \lbrack 11] Y.-K. Huang, Phys. Rev. E \textbf{83}, 036702 (2011)

\noindent \lbrack 12] S. \"{O}stlund and S. Rommer, Phys. Rev. Lett. \textbf{%
75}, 3537 (1995)

\noindent \lbrack 13] M. A. Mart\'{\i}n-Delgado and G. Sierra, Int. J. Mod.
Phys. A \textbf{11}, 3145 (1996); S. Rommer and S. \"{O}stlund, Phys. Rev. B 
\textbf{55}, 2164 (1997); I. P. McCulloch, J. Stat. Mech.: Theor. Exp.
P10014 (2007); J. I. Cirac and F. Verstraete, J. Phys. A: Math. Theor. 
\textbf{42}, 504004 (2009); U. Schollw\"{o}ck, Ann. Phys. \textbf{326}, 96
(2011)

\noindent \lbrack 14] S. R. White, Phys. Rev. B \textbf{72}, 180403 (2005)

\noindent \lbrack 15] R. Or\'{u}s and G. Vidal, Phys. Rev. B \textbf{78},
155117 (2008)

\noindent \lbrack 16] Y.-K. Huang and S.-N. Yu, Physica A \textbf{389}, 736
(2010)

\noindent \lbrack 17] L. Onsager, Phys. Rev. \textbf{65},117 (1944)

\noindent \lbrack 18] H. Au-Yang and J. H. H. Perk, Phys. Lett. A \textbf{104%
}, 131 (1984)

\end{document}